\documentstyle[epsfig]{mn}
\begin{document}
\title[MACHOS from QH Phase Transition]
{Massive Compact Halo Objects from the Relics of the
Cosmic Quark-Hadron Transition}
\author[Banerjee et al.]
{Shibaji Banerjee\thanks{Electronic Mail 
: phys@bosemain.boseinst.ac.in}\(^{a,b}\),
Abhijit Bhattacharyya\thanks{Electronic Mail 
: abhijit@theory.saha.ernet.in}\(^{c,d}\),
Sanjay K. Ghosh\thanks{Electronic Mail 
: sanjay@bosemain.boseinst.ac.in}\(^{a,e}\), 
\newauthor
Sibaji Raha\thanks{Corresponding Author}
\thanks{Electronic Mail :
sibaji@bosemain.boseinst.ac.in}\(^{a,f}\)
Bikash Sinha\thanks{Electronic Mail 
: bikash@veccal.ernet.in} \(^{c,e}\) and 
Hiroshi Toki\thanks{Electronic Mail 
: toki@rcnp.osaka-u.ac.jp}\(^{f}\)
\\
\(^a\) Department of Physics, Bose Institute,
93/1, A.P.C.Road, Kolkata 700 009, INDIA \\
\(^b\) Physics Department, St. Xavier's College, 30,
Park Street, Kolkata 700 016, INDIA \\
\(^c\) Theory Division, Saha Institute of Nuclear Physics, 1/AF,
Bidhannagar, Kolkata 700 064, INDIA \\
\(^d\) Physics Department, Scottish Church College,
3 Cornwallis Sq. Kolkata 700 006, INDIA \\
\(^e\) Variable Energy Cyclotron Centre, 1/AF, Bidhannagar, Kolkata
700 064, INDIA \\
\(^f\) Research Center for Nuclear Physics, Osaka 
University, Mihogaoka 10-1, Ibaraki City, Osaka 567-0047, JAPAN
}
\maketitle
\begin{abstract}
The existence of compact gravitational lenses, with masses around
0.5 \( M_{\odot} \), has been reported in the halo of the Milky Way. 
The nature of these dark lenses is as yet obscure, particularly 
because these objects have masses well above the threshold for nuclear 
fusion. In this work, we show that they find a natural explanation as 
being the evolutionary product of the metastable false vacuum 
domains (the so-called strange quark nuggets) formed in a first order 
cosmic quark-hadron transition.
\end{abstract}
One of the abiding mysteries in the so-called standard
cosmological model is the nature of the dark matter. It is universally
accepted that there is an abundance of matter in the universe which is
non-luminous, due to its very weak interaction, if at all, with the
other forms of matter, excepting of course the gravitational attraction.
The present consensus (for a review, see \cite{turner1,turner2}) based on 
recent experimental data is that the universe is flat and that a 
sizable amount of the dark matter is "cold", {\it i.e.} nonrelativistic,
at the time of decoupling (we are not addressing the issue of dark
energy in this work). Speculations as to the nature of dark matter 
are numerous, often bordering on the exotic, and searches for 
such exotic matter is a very active field of astroparticle physics. 
In recent years, there has been experimental evidence 
\cite{alcock1,aubourg1} for at least one form of dark
matter - the Massive Astrophysical Compact Halo Objects (MACHO) 
- detected through gravitational microlensing effects proposed by 
Paczynski \shortcite{pacz} some years ago. To date, there is 
no clear picture as to what these objects are made of. In this 
work, we show that they find a natural explanation as leftover 
relics from the {\it putative} first order cosmic quark - hadron 
phase transition.
\par
Since the first discovery of MACHO only a few years ago, a lot of effort
has been spent in studying them. Based on about 13 - 17 Milky Way
halo MACHOs detected in the direction of LMC - the Large Magellanic Cloud
(we are not considering the events found toward the galactic bulge), the
MACHOs are expected to be in the mass range (0.15-0.95) \( M_{\odot} \),
with the most probable mass being in the vicinity of 0.5 \( M_{\odot} \)
\cite{sutherland,alcock2}, substantially higher than the 
fusion threshold of 0.08 \( M_{\odot} \). The MACHO collaboration suggests 
that the lenses are in the galactic halo. Assuming that they are 
subject to the limit on the total baryon number imposed by the 
Big Bang Nucleosynthesis (BBN), there have been suggestions that 
they could be white dwarfs \cite{fields,freese}. 
It is difficult to reconcile this with the absence of sufficient 
active progenitors of appropriate masses in the galactic halo. Moreover, 
recent studies have shown that these objects are unlikely to be 
white dwarfs, even if they were as faint as blue dwarfs, since this 
will violate some of the very well known results of BBN \cite{freese}.
There have also been 
suggestions \cite{schramm,jedam1,jedam2} that they could be 
primordial black holes (PBHs) ( \( \sim \) 1 \( M_{\odot} \) ), 
arising from horizon scale fluctuations triggered by pre-existing 
density fluctuations during the cosmic quark-hadron phase transition. 
The problem with this suggestion is that the density contrast 
necessary for the formation of PBH is much larger than the 
pre-existing density contrast obtained from the common 
inflationary scenarios. The enhancement contributed by the QCD phase 
transition is not large enough for this purpose. As a result a 
fine tuning of the initial density contrast becomes essential which 
may still not be good enough to produce cosmologically relevant amount 
of PBH \cite{schmid}. Alternately, Evans, Gyuk \& Turner \shortcite{evans}
suggested that some of the lenses are stars in the Milky Way disk
which lie along the line of sight to the LMC. 
Gyuk \& Gates \shortcite{gyuk} examined a thick disk model, 
which would lower the lens mass estimate. Aubourg et al. 
\shortcite{aubourg2}  suggested that the events could arise from 
self-lensing of the LMC. Zaritsky \& Lin \shortcite{zaritsky} have argued
that the lenses are probably the evidence of a tidal tail arising from the
interaction of LMC and the Milky Way or even a LMC-SMC (Small Magellanic
Cloud) interaction. These explanations are primarily motivated by the
difficulty of reconciling the existence of MACHOs with the known 
populations of low mass stars in the galactic disks.
\par
Adopting the viewpoint that the lensing MACHOs are indeed in the Milky
Way halo, we propose that they have evolved out of the quark nuggets
which could have been formed in a first order cosmic quark - hadron phase
transition, at a temperature of  \( \sim \) 100 MeV during the microsecond era
of the early universe. The order of the deconfinement phase transition is
an unsettled issue till date. Lattice gauge theory suggests that in a pure
({\it i.e.} only gluons)  \( SU(3) \) gauge theory, the deconfinement phase
transition is of first order. In the presence of dynamical quarks on the
lattice, there is no unambiguous way to study the deconfinement transition;
one investigates the chiral transition. Although commonly treated to be
equivalent, there is no reason why, or if at all, these two phase
transitions should be simultaneous or of the same order \cite{alam1}.
The order of the chiral phase transition depends rather crucially 
on the strange quark mass. If the strange quark is heavy, then the chiral 
phase transition is probably of first order. Otherwise, it may be of 
second order. The strange quark mass being of the order of the QCD scale, 
the situation is still controversial \cite{blazo}. There are additional 
ambiguities arising from finite size effects of the lattice which may tend 
to mask the true order of the transition. Our interest here is in the 
deconfinement transition. If it is indeed of first order, the finite size 
effects which could mask it would be negligibly small in the 
early universe. In such circumstances, Witten (1984) argued, in a 
seminal paper, that strange quark matter could be the {\it true} 
ground state of {\it Quantum Chromodynamics} (QCD) and that a 
substantial amount of baryon number could be trapped in the quark phase 
which could evolve into strange quark nuggets (SQNs) through weak 
interactions. (For a brief review of the formation of SQNs, 
see Alam, Raha \& Sinha \shortcite{alam3}.) QCD - motivated studies 
of baryon evaporation from SQNs have established \cite{pijush1,sumi}
that primordial SQNs with baryon 
numbers above \(\sim\) 10\(^{40 - 42}\) would be cosmologically stable. 
We have recently shown that without much fine tuning, these 
stable SQNs could provide even the entire closure 
density (\(\Omega \sim \) 1) \cite{alam3}. Thus, the entire cold dark 
matter (CDM) ( \( \Omega_{\mathrm{CDM}} \sim \) 0.3-0.35 ) could 
easily be explained by stable SQNs.
\par
We can estimate the size of the SQNs formed in the {\it first order} 
cosmic QCD transition in the manner prescribed by Kodama, Sasaki 
and Sato \cite{kodama} in the context of the GUT phase transition. 
For the sake of brevity, let us recapitulate very briefly the salient 
points here; for details, please see Alam {\it et al} (1999) and 
Bhattacharyya {\it et al} (2000). Describing the cosmological
scale factor \( R \) and the coordinate radius \( X \) in the 
Robertson-Walker metric through the relation
\begin{eqnarray} 
ds^2 &=& -dt^2 + R^2 dx^2 \nonumber \\
&=& -dt^2 + R^2\{dX^2 + X^2(sin^2 \theta d\phi^2 
+ d\theta^2)\}, 
\end{eqnarray}
one can solve for the evolution of the scale factor \( R(t) \) in the 
mixed phase of the first order transition. In a bubble nucleation 
description of the QCD transition, hadronic matter starts to appear 
as individual bubbles in the quark-gluon phase. With progressing time, 
they expand, more and more bubbles appear, coalesce and finally, when 
a critical fraction of the total volume is occupied by the hadronic 
phase, a continuous network of hadronic bubbles form (percolation) in 
which the quark bubbles get trapped, eventually evolving to SQNs. 
The time at 
which the trapping of the false vacuum (quark phase) happens 
is the percolation time \( t_p \), whereas 
the time when the phase transition starts is denoted by \( t_i \). 
Then, the probability that a spherical\footnote{For the QCD bubbles, it 
is believed that there is a sizable surface tension which would 
facilitate spherical bubbles.} region of co-coordinate radius \( X \) 
lies entirely within the quark bubbles would obviously depend on 
the nucleation rate of the bubbles as well as the coordinate radius 
\( X(t_p,t_i) \) of bubbles which nucleated at \( t_i \) 
and grew till \( t_p \). For a nucleation rate \( I(t) \), this 
probability \( P(X,t_p) \) is given by 
\begin{equation}
P(X,t_p) = exp \left[-\frac{4 \pi}{3}
\int_{t_i}^{t_p} dt I(t) R^3(t) [X + X(t_p,t_i)]^3\right].
\end{equation}
After some algebra \cite{bhatta2}, it can be shown that if all the 
cold dark matter (CDM) 
is believed to arise from SQNs, then their size distribution peaks, for 
reasonable nucleation rates, at baryon number \( \sim \) 10\(^{42 - 44}\), 
evidently in the stable sector. It was also seen that there were almost 
no SQNs with baryon number exceeding 10\(^{46-47}\), comfortably lower 
than the horizon limit of \( \sim 10^{50} \) baryons at that time. 
Since  \(\Omega_{\mathrm{B}}\) is only about 0.04 from BBN, 
\( \Omega_{\mathrm{CDM}} \) in the form of SQNs would correspond to
\( \sim 10^{51} \) baryons so that there should be 
10\(^{7-9}\) such nuggets within the horizon limit at the microsecond 
epoch, just after the QCD phase transition \cite{alam2,alam3}. We shall 
return to this issue later on.
\par
It is therefore most relevant to investigate the fate of these SQNs.
Since the number distribution of the SQNs is sharply peaked 
\cite{bhatta2}, we shall assume, for our present purpose, that all 
the SQNs have the same baryon number. 
\par
The SQN's formed during the cosmic QCD phase transition at 
T \(\sim 100 \) MeV have high masses (\( \sim 10^{44} \)GeV) and 
sizes (\( R_N \sim 1 \)m) compared to the other particles (like 
the usual baryons or leptons) which inhabit this primeaval universe. 
These other particles cannot form structures until the temperature 
of the ambient universe falls below a certain 
critical temperature characteristic of such particles; till then, 
they remain in thermodynamic equilibrium with the radiation and 
other species of particles. This characteristic temperature is 
called the freezeout temperature for the corresponding particle.
Obviously the freezeout occurs earlier for massive particles for the
same interaction strength. In the context of 
cosmological expansion of the universe this has important 
implications ; the 'frozen' objects can form structures. These 
structures do not participate in the expansion in the sense that 
the distance between the subparts do not increase with the scale
size and only their number increases due to the cosmological scale 
factor.  
\par
For the SQN's, however, the story is especially interesting. Even if they 
continue to be in kinetic equilibrium due to the radiation pressure 
(photons and neutrinos) acting on them, their velocity would be extremely 
non relativistic. Also their mutual separation would be considerably 
larger than their radii; for example, at T \(\sim 100 \)MeV, the mutual 
separation between the SQN's (of size \( \sim 10^{44} \) baryons) is 
estimated to be around \( \sim 300 \)m. It is then obvious that the 
SQN's do not lend themselves to be treated in a hydrodynamical framework; 
they behave rather like discrete bodies in the background of the 
radiation fluid. They thus experience the radiation pressure, quite 
substantial because of their large surface area as well as the 
gravitational potential due to the other SQN's.
\par
In such a situation, one might be tempted to assume that since the SQN's 
are distributed sparsely in space and interact only feebly with the other
SQN's through gravitational interaction, they might as well 
remain forever in that state. This, in fact, is quite wrong, as we
demonstrate below.
\par
The fact that the nuggets remain almost static is hardly an issue 
which requires justification. The two kinds of motion that they can 
have are random thermal motion and the 
motion in the gravitational well provided by the other SQN's. This 
other kind of motion is typically estimated using the virial theorem, 
treating the SQN's as a system of particles moving under mutual 
gravitational interaction \cite{bhat,peebles}. The kinetic energy 
( \( K \) ) and potential
energy \( V \) of the nuggets at temperature T = 100 MeV can be 
estimated as, 
\begin{eqnarray}
K= \frac{3}{2} N k_{b} T  \nonumber \\
V=\sum_{i,j} G\frac{M_{i} M_{j}}{R_{i,j}}=
\frac{GM^{2} N^{2}}{2 R_{av}} 
\end{eqnarray}
where \( k_b \) is the Boltzmann constant, \( M_{i}, M_{j} \) are 
the masses of the ith and jth nugget, \( R_{i,j} \) is the distance 
between them and \( R_{av} \) is the average inter-nugget distance. 
Substituting the number of nuggets \( N = 10^7 \), the baryon number of 
each nugget to be \( 10^{44} \) and \( R_{av} = 300 m \), one gets 
\( K = 2.4 \times 10^{-4} \) and \( V = 3.09 \times 10^{35} \) 
(in MKS units) so that the ratio of \( K \) and \( \frac{V}{2} \) 
becomes \( \sim 10^{-39} \). Thus it is impossible for these 
objects to form stable systems, orbiting round each other. On the other
hand the smallness of the kinetic energy shows that gravitational 
collapse might be a possible fate. 
\par
Such, of course, would not be the case for any other massive 
particles like baryons; their masses being much smaller than SQN, 
the kinetic energy would continue to be very large 
till very low temperatures. More seriously, the Virial theorem can be
applied only to systems whose motion is sustained. 
For SQNs, a notable property is that they become more and more bound 
if they grow in size. Thus SQNs would absorb baryons impinging on
them and grow in size. Also, if two SQNs collide, they would naturally 
tend to merge. In all such cases, they would lose kinetic energy,
making the Virial theorem inapplicable. 
\par
One can argue that the mutual interaction between uniformly 
dispersed particles would prevent these particles from forming a 
collapsed structure, but that argument holds only in a static and 
infinite universe, which we know our universe is not. Also a perfectly
uniform distribution of discrete bodies is an unrealistic idealization 
and there must exist some net gravitational attraction on each SQN.
The only agent that can prevent a collapse under this gravitational pull 
is the radiation pressure, and indeed its effect remains 
quite substantial until the drop in the temperature of the ambient 
universe weakens the radiation pressure below a certain critical value. 
In what follows, we try to obtain an estimate for the point of time at 
which this can happen. 
\par
It should be mentioned at this juncture that for the system of discrete 
SQN's suspended in the radiation fluid, a detailed numerical simulation 
would be essential before any definite conclusion about their temporal 
evolution can be arrived at. This is a quite involved problem, 
especially since the number of SQN's within the event horizon, as 
also their mutual separation, keeps increasing with time. Our 
purpose in the present work is to examine whether such an effort 
would indeed be justified.  
\par
Let us now consider the possibility of two nuggets coalescing together 
under gravity, overcoming the radiation pressure. The mean separation 
of these nuggets and hence their gravitational interaction are 
determined by the temperature of the universe. If the entire CDM
comes from SQNs, the total baryon number contained in 
them within the horizon at the QCD transition temperature 
( \( \sim \) 100 MeV) would be  \( \sim 10^{51} \) (see above). 
For SQNs of baryon number \( b_{N} \) each, the number of SQNs 
within the horizon at that time would be just 
\( \left( 10^{51} /b_{N}\right) \). 
Now, in the radiation dominated era the temperature dependence of density 
\( n_{N} \sim T^3 \), horizon volume \( V_{H} \) varies with time as 
\( t^3 \), { \it{i.e.} } \( V _{H} \sim T^{-6} \) and hence the 
variation of the total number inside the horizon volume will be 
\(N_{N} \sim T^{-3} \). So at any later time, the number of SQNs within 
the horizon ( \( N_{N} \) ) and their density ( \(n_{N}\) ) as a 
function of temperature would be given by : 
\begin{eqnarray}
N_{N}(T) & \cong & {\frac{ 10^{51}} {b_{N}}}
\left(\frac{100\mathrm{MeV}}{\mathrm{T}}\right)^3 \label{Nn} \\
n_{N}(T) & = & \frac{N_{N}}{V_{H}}=
\frac{3N_{N}}{4 \pi (2t)^{3}}\label{nn} 
\end{eqnarray}
where the time \( t \) and the temperature \( T \) are related in the 
radiation dominated era by the relation :
\begin{equation}
t = 0.3 g_{*}^{-1/2}\frac {m_{pl}}{T^{2}} \label{tT}
\end{equation}
with \( g_{*} \) being \( \sim \) 17.25 after the QCD transition 
(Alam et al. 1999).
\par
From the above, it is obvious that the density of SQNs decreases as \( 
t^{-3/2} \) so that their mutual separation increases as \( t^{1/2} 
\). Therefore, the force of their mutual gravitational pull will 
decrease as \( t^{-1} \). On the other hand, the force due to the 
radiation pressure (photons and neutrinos) resisting motion under 
gravity would be proportional to the radiation energy density, which 
decreases as \( T^{4} \) or \( t^{-2} \). It is thus reasonable to 
expect that at some time, not too distant, the gravitational pull 
would win over the radiation pressure, causing the SQNs to coalesce 
under their mutual gravitational pull.
The expression for the gravitational force as a function of temperature
T can written as :
\begin{equation}
F_{\mathrm{grav}}=\frac {G {b_N}^2 {m_n}^2} {{\bar{r}_{nn}(T)}^2}
\end{equation}
where \( b_N \) is the baryon number of each SQN and \( m_n \) is the 
baryon mass. \( \bar{r}_{nn}(T) \) is the mean separation between 
two nuggets and is given by the cube root of the ratio \( \kappa \) of 
total volume available and the total number of nuggets
\begin{equation}
\kappa = \frac {1.114 \times 10^{-12}c^3}{T^3}
\end{equation}
The force due to the radiation pressure on the nuggets may be roughly
estimated as follows. We consider two objects (of the size of a typical SQN) 
approaching each other due to gravitational interaction, overcoming the 
resistance due to the radiation pressure. The usual isotropic radiation 
pressure is \(  \frac{1}{3} \rho c^2 \), where \( \rho \) is the total
energy density, including all relativistic species. The nuggets will have 
to overcome an additional pressure resisting their mutual motion, which is
given by \( \frac{1}{3} \rho c^2 (\gamma - 1) \); the additional pressure 
arises from a compression of the radiation fluid due to the motion of the 
SQN. The moving SQN would become a prolate ellipsoid (with its minor axis 
in the direction of motion due to Lorentz contraction), whose surface area 
is given by 
\(2 \pi R_N^2 (1 + \frac {sin^{-1} \epsilon } {\gamma \epsilon}) \). The
eccentricity \( \epsilon \) is related to the Lorentz factor \( \gamma \) 
as \( \epsilon = \frac{\sqrt{\gamma^2 -1}} {\gamma} \). For small values of
\( \epsilon \) (small \( \gamma) \), \( sin^{-1} \epsilon \sim \epsilon \), so
that the surface area becomes \(2 \pi R_N^2 \frac{\gamma + 1} { \gamma} \). 
Thus the total radiation force resisting the motion of SQNs is
\begin{equation}
F_{\mathrm{rad}}=\frac{1}{3} \rho_{\mathrm{rad}} c v_{\mathrm{fall}}
(\pi R_N^2) \beta \gamma
\end{equation}
where \( \rho_{\mathrm{rad}} \) is the total energy density at 
temperature \( T \), \( v_{\mathrm{fall}} \) or \(\beta c\) is the
velocity of SQNs determined by mutual gravitational field and \( \gamma \) 
is \( 1 / \sqrt{1-\beta^2} \). The quantities  \( F_{\mathrm{rad}} \),
\( \beta \) and \( \gamma \) all depend on the temperature of the epoch
under consideration. (It is worth mentioning at this point that the \( 
t \) dependence of \( F_{\mathrm{rad}} \) is actually \( t^{-5/2} \), 
sharper than the \( t^{-2} \) estimated above, because of the \( 
v_{\mathrm{fall}} \), which goes as \( t^{-1/4} \).)
The ratio of these two forces is plotted against temperature in 
figure 1 for two SQNs with initial baryon number \( 10^{42} \) each. 
It is obvious 
from the figure that ratio \( F_{\mathrm{grav}}/F_{\mathrm{rad}} \)
is very small initially. As a result, the nuggets will remain 
separated due to the radiation pressure. For temperatures lower than 
a critical value \( T_{\mathrm{cl}} \), the gravitational force starts 
dominating, facilitating the coalescence of the SQNs under mutual 
gravity.
\par
Let us now estimate the mass of the clumped SQNs, assuming that 
all of them within the horizon at the critical temperature will 
coalesce together. This is in fact a conservative estimate, since the
SQNs, although starting to move toward one another at 
\( T_{\mathrm{cl}} \), will take a finite time to actually coalesce, 
during which interval more SQNs will arrive within the horizon.
\par
In table 1, we show the values of \( T_{\mathrm{cl}} \) for SQNs 
of different initial baryon numbers along with the final masses of 
the clumped SQNs under the conservative assumption mentioned above.
\begin{figure} 
\epsfig{file=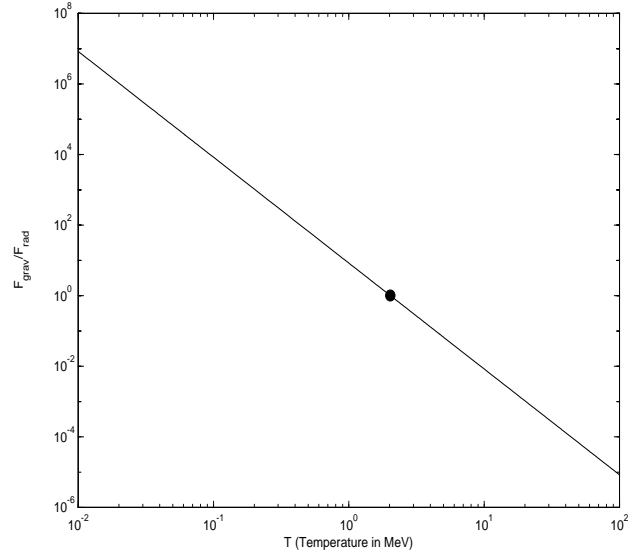,height=6in,width=3.6in}
\vskip -1in
\caption{Variation of the ratio 
\( F_{\mathrm{grav}} / F_{\mathrm{rad}} \)
with temperature. The dot represents the point 
where the ratio assumes the value 1.}
\label{fig1}
\end{figure}
\begin{table}
\label{tab1}
\caption{Critical temperatures ( \( T_{\mathrm{cl}}\) ) of SQNs of 
different initial sizes \( b_{N} \), the total number \( N_{N} \) 
of SQNs that coalesce together and their total final mass in 
solar mass units.}
\begin{tabular}{|c|c|c|c|}
\( b_{N} \) &\( T_{\mathrm{cl}} \) (MeV) & \( N_{N} \) &\( M/{M\odot} \) \\
\hline
& & & \\  
\( 10^{42} \) & 1.6  & \( 2.44 \times 10^{14} \) & 0.24  \\
& & & \\
\( 10^{44} \) & 4.45 & \( 1.13 \times 10^{11} \) & 0.01  \\
& & & \\
\( 10^{46} \) & 20.6 & \( 1.1 \times 10^7 \)   & 0.0001 \\
& & & \\
\hline
\end{tabular} 
\end{table}
\par
It is obvious that there can be no further clumping of these already
clumped SQNs; the density of such objects would be  too small within the
horizon for further clumping. Thus these objects would
survive till today and perhaps manifest themselves as MACHOs. 
It is to be reiterated that the masses of the clumped SQN's given in 
table 1 are the lower limits and the final masses of these 
MACHO candidates will be larger. (The case for \( b_{N} = 10^{46} \) 
is not of much interest, especially since such high values of 
\( b_{N} \) are unlikely for the reasonable nucleation rates 
\cite{bhatta1,bhatta2}; we therefore restrict ourselves to the other 
cases in table 1 in what follows.) 
A more detailed estimate of the masses will require a
detailed simulation, but very preliminary estimates indicate that 
they could be 2-3 times bigger than the values quoted in table 1.
\par
The total number of such clumped SQNs ( \( N_{\mathrm{macho}} \) ) 
within the horizon today is evaluated in the following way. With the
temperature \( \sim 3 {^0}K \) and time \( \sim 4 \times 10^{17} \)
seconds, the total amount of visible baryons within the horizon volume 
can be evaluated using photon to baryon 
ratio \( \eta \sim  10^{-10} \). The amount of baryons in the CDM will
be \( \frac{\Omega_{\mathrm{CDM}}}{\Omega_{\mathrm{B}}} \) times the
total number of visible baryons. This comes out to be \( \sim 1.6 \times
10^{79} \), \( \Omega_{\mathrm{CDM}} \) and \( \Omega_{\mathrm{B}} \)
being 0.3 and 0.01 respectively. The total number of baryons in a MACHO
is \( b_{N} \times N_{N} \) i.e. \( 2.44 \times 10^{56} \) and
\( 1.13 \times 10^{55} \) for initial nugget sizes \( 10^{42} \) and
\( 10^{44} \) respectively. The quantities \( b_{N} \) and
\( N_{N} \) are taken from the Table 1. So dividing the total number 
of baryons in CDM by that in a MACHO,
the \( N_{\mathrm{macho}} \) comes out to be in the range 
\( \sim 10^{23-24} \). 
\par
We can also mention here that 
if the MACHOs are indeed made up of quark matter, then they cannot 
grow to arbitrarily large sizes. Within the (phenomenological) Bag model 
picture \cite{chodos} of QCD confinement, where a constant vacuum 
energy density (called the Bag constant) in a cavity containing 
the quarks serves to keep them confined within the cavity, we have 
earlier investigated \cite{ban} the upper limit on the mass of 
astrophysical compact quark matter objects. It was found that for a 
canonical Bag constant {\bf B} of (145 MeV) \(^4 \), this limit 
comes out to be 1.4 \( M_\odot \). The 
collapsed SQNs are safely below this limit. (It should be remarked here 
that although the value of {\bf B} in the original MIT bag model is taken 
to be {\bf B} \(^{1/4} \) = 145 MeV from the low mass
hadronic spectrum, there exist other variants of the Bag model 
\cite{hasen} , where higher values of {\bf B} are required. Even 
for {\bf B} \(^{1/4} \) = 245 MeV, this limit comes down 
to 0.54  \( M_\odot \) (Banerjee et al. 2000), which would still 
admit such SQN.
\par
As a consistency check, we can perform a theoretical estimate of the
abundance of such MACHO's in the galactic halo which is 
conventionally given by the optical 
depth. The optical depth is the probability that at any instant of time
a given star is within an angle \( \theta_{E} \) of a lens, the lens 
being the massive body (in our case MACHO) which causes the deflection 
of light. In other words, optical depth is the integral over the number 
density of lenses times the area enclosed by the Einstein ring of 
each lens. The expression for optical depth can be written
as \cite{narayan}:
\begin{equation}
\tau = \frac{4 \pi G}{c^2} D_{s}^{2}\int \rho(x) x (1-x) dx 
\end{equation}
where \( D_s \) is the distance between the observer and the source,
\( G \) is the gravitational constant and \( x=D_{d} {D_{s}}^{-1} \),
\( D_d \) being the distance between the observer and the lens. In
particular \( \rho \) is the mass-density of the MACHOs, which is of the
form \( \rho = \rho_0 \frac{1}{r^2} \) in the naive spherical halo model,
which we have adopted in our calculations. In the present case 
\( \rho_0 \) is given by 
\begin{equation}
\rho_0 = \frac{ M_{\mathrm{macho}} \times N_{\mathrm{macho}}}{4\pi R } 
\end{equation}
where 
\( R= \sqrt{D_{e}^2 + D_{s}^2 + 2 D_{e} D_{s} \cos\phi} \), 
\( \phi \) and \( D_{e} \) being the inclination of the LMC and the
distance of observer (earth) from the Galactic centre respectively. 
\( M_{\mathrm{macho}} \) 
and \( N_{\mathrm{macho}} \) are the mass of a MACHO and 
the total number of MACHOs in the Milky Way halo.  
\par
The total visible mass of the Milky Way 
( \( \sim  1.6 \times 10^{11} M_{\odot} \) ) 
corresponds to \( \sim 2\times 10^{68} \) baryons. 
This corresponds to  a factor of  \( \sim 2 \times 10^{-9} \) of all the 
visible baryons 
within the present horizon. Scaling the number of clumped SQNs within the 
horizon by the same factor yields a total number of  MACHOs,
\( N_{\mathrm{macho}} \sim 10^{13-14} \) in the Milky Way halo 
for the range of  baryon number of 
initial nuggets \( b_{N} = 10^{42-44} \). The value of  
\( D_{e} \) and \( D_{s} \) are taken to be 10 and 50 kpc, respectively.
The value of the inclination angle used here is 40 degrees. Using these 
values for a naive inverse square spherical model comprising such 
objects upto the LMC, we obtain an optical depth of  
\( \sim 10^{-6}-10^{-7} \).  
The uncertainty in this value is mainly governed by the value of  
\( \eta \), \( \Omega_{\mathrm{CDM}} \) and \( \Omega_{\mathrm{B}} \), 
and to a lesser 
extent by the specific halo model. This value compares reasonably well 
with the observed value and may be taken as a measure of reliability in 
the proposed model.
\par
As an interesting corollary, let us mention that the scenario presented
here could have other important astrophysical significance. The origin of
cosmic rays of ultra-high energy  \( \ge 10^{20} \) eV continues to be a 
puzzle. One of the proposed mechanisms \cite{pijush2} envisages a 
top-down scenario which does not require an acceleration mechanism and 
could indeed originate within our galactic halo. For our picture, such 
situations could easily arise from the merger of two or more such MACHOs, 
which would shed the extra matter so as to remain within the upper mass 
limit mentioned above. This is currently under active investigation.
\par
We thus conclude that gravitational clumping of the primordial SQNs
formed in a first order cosmic quark - hadron phase transition appears to
be a plausible and natural explanation for the observed halo MACHOs. It
is quite remarkable that we obtain quantitative agreement with the
experimental values without having to introduce any adjustable parameters
or any fine-tuning whatsoever. We may finish by quoting a famous teaching 
of John Archibald Wheeler, ``One should never do a calculation unless one 
knows an answer". Our attempt in this work has been to find an answer so 
that a calculation (in this case, a detailed simulation of the collapsing 
SQNs) can be embarked upon.
\par
The work of SB was supported in part by the Council of Scientific 
and Industrial Research, Govt. of India. SR would like to thank
the Nuclear Theory group at Brookhaven National Laboratory, 
USA, where this work was initiated, for their warm hospitality.

\end{document}